\documentclass[prb,letterpaper,twocolumn,showpacs,superscriptaddress,amsmath,amsfonts,amssymb,preprintnumbers,citesort,citeautoscript]{revtex4}
\pdfoutput=1
\usepackage[T1]{fontenc}\usepackage[latin1]{inputenc}
\usepackage{dcolumn,graphicx,color,booktabs,microtype,afterpage,collapse} 
\usepackage[charter]{mathdesign}
\renewcommand{\figurename}{Fig.}
\renewcommand{\tablename}{Table}
\makeatletter\renewcommand{\fnum@figure}[1]{\figurename~\thefigure~(color online).}\makeatother
\makeatletter\renewcommand{\fnum@table}[1]{\tablename~\thetable.}\makeatother
\definecolor{MyRed}{rgb}{0.8,0,0}
\definecolor{MyGreen}{rgb}{0.25,0.5,0.5}

\usepackage[colorlinks,plainpages=false,linkcolor=blue,urlcolor=blue,citecolor=black,pdfpagemode=UseNone,pdfstartview=FitBH]{hyperref}

\begin{document} \pagestyle{plain}

\title{Suppression of the structural phase transition and lattice softening\\in slightly underdoped Ba$_{1-x}$K$_x$Fe$_2$As$_2$ with electronic phase separation}

\author{D.\,S.\,Inosov}
\affiliation{Max-Planck-Institut für Festkörperforschung, Heisenbergstraße 1, 70569 Stuttgart, Germany}

\author{A.\,Leineweber}
\affiliation{Max-Planck-Institut für Metallforschung, Heisenbergstraße 1, 70569 Stuttgart, Germany}

\author{Xiaoping Yang}
\affiliation{Max-Planck-Institut für Festkörperforschung, Heisenbergstraße 1, 70569 Stuttgart, Germany}

\author{J.\,T.~Park}
\affiliation{Max-Planck-Institut für Festkörperforschung, Heisenbergstraße 1, 70569 Stuttgart, Germany}

\author{N.\,B.\,Christensen}
\affiliation{Laboratory for Neutron Scattering, ETHZ \& PSI, CH-5232 Villigen PSI, Switzerland}
\affiliation{Materials Research Division, Risø DTU, Technical University of Denmark, DK-4000 Roskilde, Denmark}
\affiliation{Nano-Science Center, Niels Bohr Institute, University of Copenhagen, DK-2100 Copenhagen, Denmark}

\author{R.\,Dinnebier}
\affiliation{Max-Planck-Institut für Festkörperforschung, Heisenbergstraße 1, 70569 Stuttgart, Germany}

\author{G.\,L.\,Sun}
\affiliation{Max-Planck-Institut für Festkörperforschung, Heisenbergstraße 1, 70569 Stuttgart, Germany}

\author{Ch.\,Niedermayer}
\affiliation{Laboratory for Neutron Scattering, ETHZ \& PSI, CH-5232 Villigen PSI, Switzerland}

\author{D.\,Haug}
\affiliation{Max-Planck-Institut für Festkörperforschung, Heisenbergstraße 1, 70569 Stuttgart, Germany}

\author{P.~W.~Stephens}
\affiliation{Department of Physics and Astronomy, State University of New York, Stony Brook, New York 11974-3800, USA}

\author{J.\,Stahn}
\affiliation{Laboratory for Neutron Scattering, ETHZ \& PSI, CH-5232 Villigen PSI, Switzerland}

\author{C.\,T.~Lin}
\affiliation{Max-Planck-Institut für Festkörperforschung, Heisenbergstraße 1, 70569 Stuttgart, Germany}

\author{O.\,K.\,Andersen}
\affiliation{Max-Planck-Institut für Festkörperforschung, Heisenbergstraße 1, 70569 Stuttgart, Germany}

\author{B.\,Keimer}
\affiliation{Max-Planck-Institut für Festkörperforschung, Heisenbergstraße 1, 70569 Stuttgart, Germany}

\author{V.~Hinkov}\email[Corresponding author: \vspace{4pt}]{v.hinkov@fkf.mpg.de}
\affiliation{Max-Planck-Institut für Festkörperforschung, Heisenbergstraße 1, 70569 Stuttgart, Germany}

\begin{abstract}
\noindent We present x-ray powder diffraction (XRPD) and neutron diffraction measurements on the slightly underdoped iron pnictide superconductor Ba$_{1-x}$K$_x$Fe$_2$As$_2$, $T_{\rm c}=32$\,K. Below the magnetic transition temperature $T_{\rm m}=70$\,K, both techniques show an additional broadening of the nuclear Bragg peaks, suggesting a weak structural phase transition. However, macroscopically the system does not break its tetragonal symmetry down to 15\,K. Instead, XRPD patterns at low temperature reveal an increase of the anisotropic microstrain proportionally in all directions. We associate this effect with the electronic phase separation, previously observed in the same material, and with the effect of lattice softening below the magnetic phase transition. We employ density functional theory to evaluate the distribution of atomic positions in the presence of dopant atoms both in the normal and magnetic states, and to quantify the lattice softening, showing that it can account for a major part of the observed increase of the microstrain.
\end{abstract}


\maketitle

\vspace{-5pt}\section{Introduction}\vspace{-5pt}

\noindent The recent discovery of superconductivity (SC) in layered iron arsenides \cite{KamiharaWatanabe08, WenMu08, ChenLi08, TakahashiIgawa08, ChenWu08, KitoEisaki08, ZhiAnWei08} served as a powerful impetus in the search for novel superconductors with high critical temperatures. Within this new family of compounds, the record holders for the highest known $T_{\rm c}$ are the electron-doped 1111-compounds Gd$_{1-x}$Th$_x$FeAsO \cite{WangLi08} and Sr$_{1-x}$Sm$_x$FeAsF \cite{WuXie09} (both with optimal $T_{\rm c}=56$~K), whereas among the so-called 122-compounds the highest $T_{\rm c}$ of 38\,K was reached in the hole-doped Ba$_{1-x}$K$_x$Fe$_2$As$_2$ near the optimal doping of $x\approx0.5$ \cite{RotterTegel08, WuLiu08}. The parent compounds ($x=0$) of both types of arsenides order antiferromagnetically (AFM) below a spin-density-wave (SDW) transition at temperatures $T_{\rm m}$ in the range between 140 and 200\,K, as seen by neutron scattering \cite{CruzHuang08, ZhaoHuang08, ZhaoHuang08PRB, HuangZhao08, HuangQiu08, ChenRen09, GoldmanArgyriou08, SuLink09} and local-probe methods, such as $\mu$SR \cite{AczelBaggio08, GokoAczel08, KlaussLuetkens08, JescheCaroca08} and $^{57}$Fe Mössbauer spectroscopy \cite{RotterTegel08PRB, KlaussLuetkens08}.  At low doping levels, this SDW transition is always accompanied by a structural phase transition at $T_{\rm s}$\,$\gtrsim$\,$T_{\rm m}$ from a high-temperature tetragonal (T) to a low-temperature orthorhombic or monoclinic structure, which manifests itself as a longitudinal splitting of the in-plane nuclear Bragg peaks $(hh0)\,_{\rm T}$ both in neutron scattering and x-ray diffraction experiments \cite{CruzHuang08, ZhaoHuang08, ZhaoHuang08PRB, HuangZhao08, HuangQiu08, ChenRen09, RotterTegel08PRB, JescheCaroca08, GoldmanArgyriou08, YanKreyssig08, NiBudko08, NiNandi08}. Up to now, to the best of our knowledge, there have been no reports of any iron pnictides, neither among parent nor doped compounds, where magnetic order would be observed without the development of a structural distortion.

\begin{figure}[b]\vspace{-1pt}
\includegraphics[width=\columnwidth]{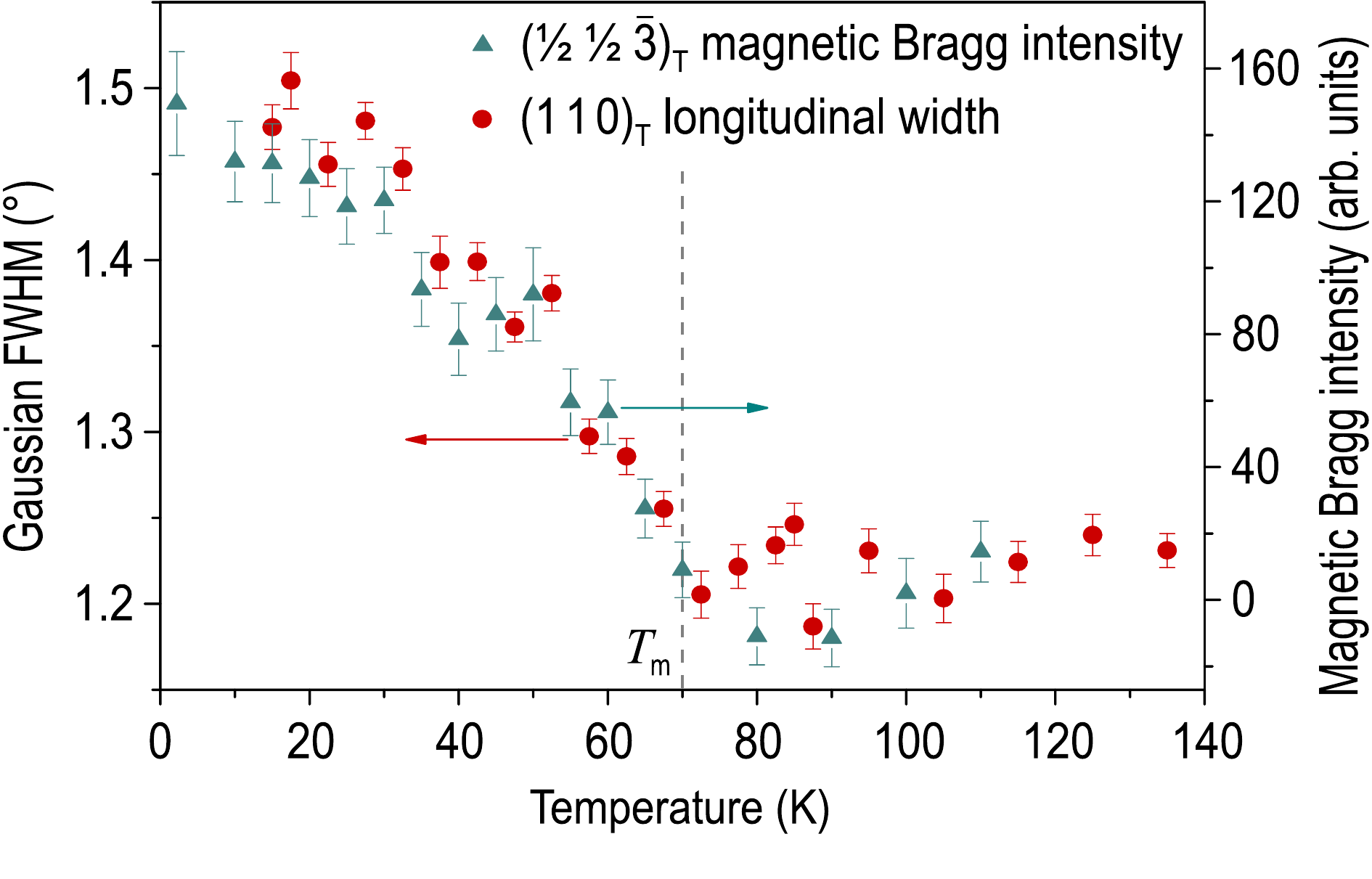}\vspace{-8pt}
\caption{Neutron diffraction data showing the temperature-dependent longitudinal broadening of the (1\,1\,0)$_{\rm T}$ nuclear Bragg reflection (\textcolor{MyRed}{$\bullet$}) overlayed with the intensity of the $\bigl(\frac{1}{\text{\protect\raisebox{1pt}{2}}}\,\frac{1}{\text{\protect\raisebox{1pt}{2}}}\,\bar{3}\bigr)_{\rm T}$ magnetic Bragg peak (\textcolor{MyGreen}{\raise0.1em\hbox{$\scriptstyle\blacktriangle$}}).\vspace{-6pt}}
\label{fig:broadening}
\end{figure}

Though it is commonly acknowledged that the magnetic and structural order parameters in iron pnictides are intimately coupled, the details of the relationship between the two phase transitions still remain a puzzle. On the one hand, in 1111-compounds (but not in 122-compounds \cite{HuangQiu08}) the structural phase transition precedes the magnetic one \cite{CruzHuang08, ZhaoHuang08}, suggesting itself as the driving force for the magnetic anisotropy of the SDW phase. On the other hand, the experimentally observed structural distortion cannot be reproduced in non-magnetic calculations \cite{MazinJohannes09, MazinJohannesBoeri08}. Therefore most theories consider the SDW instability an intrinsic property of the electronic system, driven either by the nesting of the electron- and hole-like Fermi surface sheets \cite{ChubukovEfremov08, KorshunovEremin08, MazinSingh08, DongZhang08, KurokiOnari08, YareskoLiu08} or by the local superexchange interactions in the framework of the Heisenberg model \cite{Yildirim08, SiAbrahams08, MaLu08, FangYao08, XuMuller08}. Both scenarios imply that the structural phase transition occurs as a consequence of the AFM ordering, and its somewhat higher transition temperature is explained as a response to anisotropic AFM fluctuations that persist even above $T_{\rm m}$ \cite{FangYao08, XuMuller08, MazinJohannes09, ZabolotnyyInosov09}.

In this paper, we combine neutron scattering and x-ray powder diffraction (XRPD) experiments, along with theoretical calculations, to study the interplay between the magnetic and structural phase transitions in a slightly underdoped 122-compound Ba$_{1-x}$K$_x$Fe$_2$As$_2$ (BKFA), $T_{\rm c}=32$\,K, in which the onset of a phase-separated magnetic order occurs at $T_{\rm m}=70$\,K according to our recent study performed on the same samples \cite{ParkInosov09}. Our experimental evidence indicates that macroscopically the sample preserves its tetragonal symmetry down to 15\,K, well below $T_{\rm m}$. Instead of the structural transition to an orthorhombic phase at low temperatures, seen in more underdoped BKFA samples \cite{ChenRen09, RotterTegel08PRB}, here the lattice reacts to the magnetic order only microscopically, by an increase of the microstrain as observed in our XRPD measurements, without a macroscopic breakdown of the lattice symmetry. We argue that such an effect is most probably related to a softening of the lattice below the magnetic phase transition in comparison to the high-temperature non-magnetic state, whereas the phase-separated coexistence of twinned magnetic domains and the non-magnetic phase \cite{ParkInosov09} suppresses the structural phase transition beyond the experimentally detectable limit, in spite of a relatively high SDW transition temperature.

\vspace{-5pt}\section{Sample preparation}\vspace{-5pt}

The single crystals of BKFA used for the present study were grown using Sn as flux in a zirconia crucible sealed in a quartz ampoule filled with Ar. A mixture of Ba, K, Fe, As, and Sn in a wt. ratio of BKFA:Sn = 1:85 was heated in a box furnace up to 850$^\circ$C and kept constant for 2\,--\,4 hours to soak the sample in a homogeneous melt. The cooling rate of 3$^\circ$C/h was then applied to decrease the temperature to 550$^\circ$C, and the grown crystals were then decanted from the flux \cite{SunSunLin08}. Sample characterization by resistivity and dc susceptibility measurements \cite{ParkInosov09} revealed a sharp SC transition at $T_{\rm c,\,onset}=(32\pm1)$~K, reproducible among different samples from the same batch. The same samples have been extensively studied by angle-resolved photoelectron spectroscopy (ARPES) \cite{ZabolotnyyInosov09, EvtushinskyInosov09, EvtushinskyInosovNJP} and muon-spin rotation ($\mu$SR) \cite{ParkInosov09, KhasanovAmato09}.

The neutron diffraction measurements were done on a $\sim$\,30\,mg single crystal with in-plane ($hh0$) and out-of-plane ($00l$) mosaicities better than 1.5° and 2.5°, respectively, as determined from the full width at half maximum (FWHM) of the rocking curves. A few smaller single crystals from the same batch were ground into powder for XRPD analysis. The sample was then prepared by sprinkling a small amount of the powder onto a flat brass sample holder.

\begin{figure}[t]
\includegraphics[width=\columnwidth]{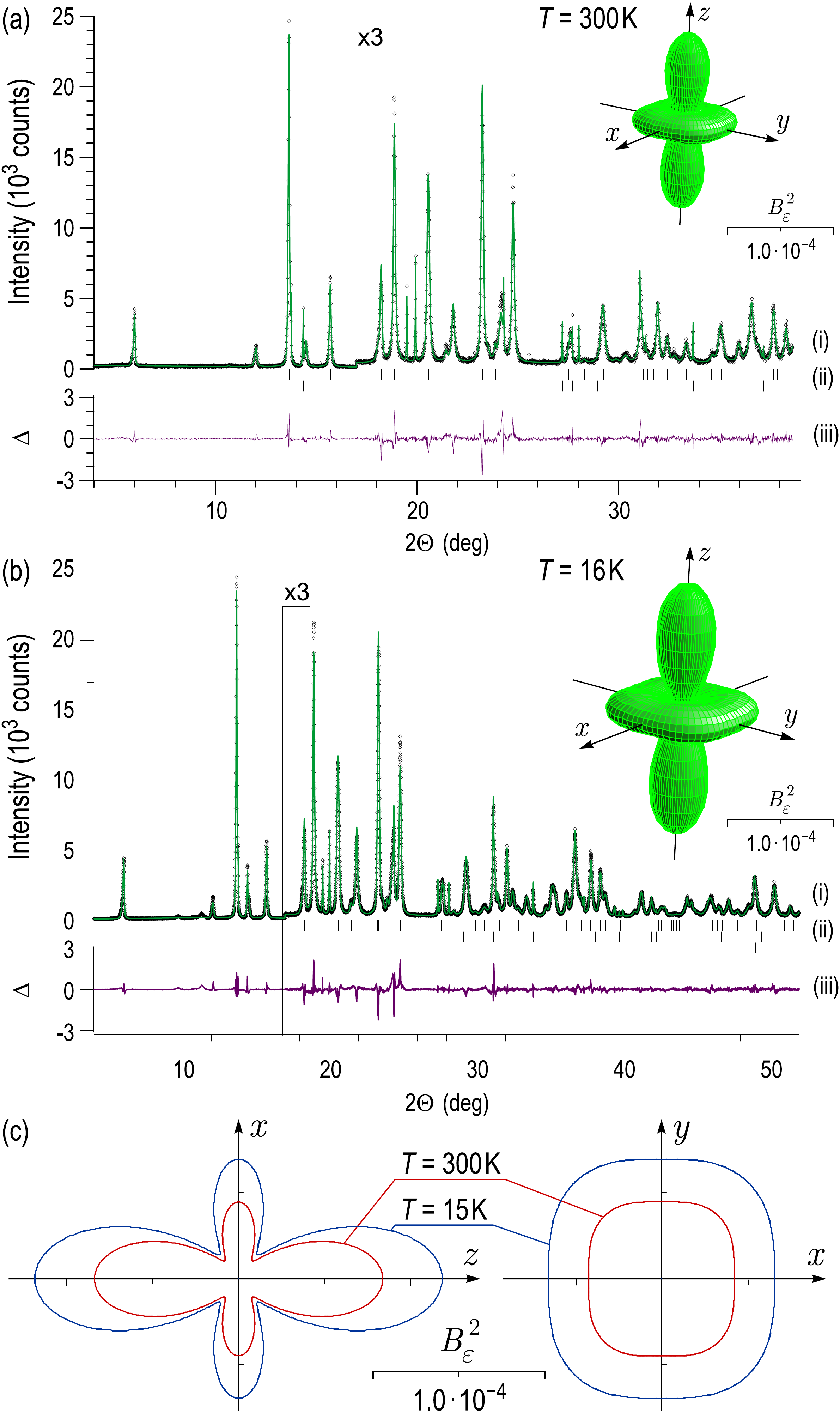}
\caption{Panels (a) and (b) present XRPD data measured at 300\,K and 15\,K, respectively. (i) Scattered x-ray intensity as a function of the diffraction angle $2\Theta$ ($\lambda=0.7$\,\AA) fitted to the tetra\-gonal \textit{I}4/\textit{mmm} space group. For $2\Theta>17^\circ$ the plots are enlarged by a factor of three. The fit includes a few wt.\,\% of tetra\-gonal $\beta$-tin from the flux as an impurity phase and some reflections of the brass sample holder as indicated by the reflection markers in (ii). (iii)~The difference $\Delta$ between the ex\-pe\-rimental points and the fitting curve. The insets show tensor surfaces representing the normalized anisotropic microstrain distribution along different crystallographic directions. The distance of the surface from the origin corresponds the squared full width at half maximum (FWHM) of the microstrain $B_\varepsilon^2$ along the corresponding directions in real space. The $x$-$z$ and $x$-\kern-0.5pt$y$ cross-sections of both surfaces are shown in panel (c) for comparison.}
\label{fig:xray}
\end{figure}

\vspace{-5pt}\section{Neutron diffraction}\vspace{-5pt}

We have measured the longitudinal width of the (1\,1\,0)$_{\rm T}$ nuclear Bragg reflection as a function of temperature, which is plotted in Fig.\,\ref{fig:broadening} together with the intensity of the $\bigl(\frac{1}{\text{\protect\raisebox{1pt}{2}}}\,\frac{1}{\text{\protect\raisebox{1pt}{2}}}\,\bar{3}\bigr)_{\rm T}$ magnetic Bragg peak. One can clearly see the broadening of the nuclear Bragg peak at low temperatures, with an onset at $T_{\rm m}$, which perfectly follows the magnetic intensity, and amounts to $\sim$\,20\% as $T\rightarrow0$. The most straightforward explanation for such broadening would be a weak orthorhombic distortion that leads to a splitting of the peak that is masked by the experimental resolution, as was also previously observed whenever the AFM order was suppressed either by doping, as in CeFeAsO$_{0.94}$F$_{0.06}$ at low temperature [Ref.\,\onlinecite{ZhaoHuang08}, Fig.\,2\,(d)], or by temperature, as in the parent compound LaFeAsO at $T=138$\,K [Ref.\,\onlinecite{CruzHuang08}, Fig.\,4\,(inset)].

To check this interpretation, we have performed XRPD measurements of the same samples, with subsequent ana\-lysis of the microstrain anisotropy, which is known to be helpful in detecting minute structural distortions related to possible phase transitions \cite{Leineweber06, Leineweber07}.

\vspace{-5pt}\section{X-ray powder diffraction}\vspace{-5pt}

The XRPD data for the structure refinement were collected at room temperature and at 15\,K, as shown in Fig.\,\ref{fig:xray} (a) and (b). The sample was placed in a closed cycle cryostat. X-rays of 0.7\,\AA\ wavelength were selected by a double Si(111) monochromator. The wavelength and zero-point error were calibrated using 8 precisely measured peaks of the NBS1976 flat plate alumina standard. The diffracted beam was analyzed by reflection from a Ge(111) crystal before a NaI scintillation detector. Data were taken at each $2\Theta$ step of 0.005° from 3° to 38.6° at room temperature and 2° to 52° at 15\,K. The sample was rocked during the measurement for better particle statistics. All data were normalized for storage ring current decay by an ionization chamber monitor.

XRPD data were analyzed using the program TOPAS (\textit{Bruker-AXS}). Both high- and low-temperature data could be interpreted in terms of a tetragonal \textit{I}4/\textit{mmm} space group symmetry both at room temperature and at $T=15$~K (see Fig.\,\ref{fig:xray}). As impurity phases, a few wt.\,\% of tetragonal $\beta$-tin from the flux and some reflections of the brass sample holder were included in the refinement. The analysis of the anisotropic peak broadening in the powder pattern due to a microstrain distribution was performed using the Cartesian parametrization by Leineweber \cite{Leineweber06, Leineweber07}.

The lattice parameters of the sample, as determined from XRPD by Rietveld refinement using the fundamental parameters approach of TOPAS \cite{ChearyCoelho05}, are $a$\,=\,$b$\,=\,3.9111(1)\,\AA\ and $c$\,=\,13.3392(6)\,\AA\ at room temperature and $a$\,=\,$b$\,=\,3.90075(7)\,\AA\ and $c$\,=\,13.2476(3)\,\AA\ at 15\,K, which corresponds to a 1.2\% decrease in the unit cell volume at low temperature. From the dependence of the lattice parameters on doping \cite{RotterPangerl08}, the average potassium content of $x=0.4$ could be determined, in agreement with the results of our energy-dispersive x-ray analysis. No evidence was found for an orthorhombic distortion of the tetragonal lattice at low temperature. This conclusion is based on the absence of any orthorhombic splitting of the Bragg reflections and the refinement of the lattice parameters. The isotropic microstrain distribution in the $(hk0)$ plane also does not hint at an orthorhombic distortion.

\begin{figure}[b]
\includegraphics[width=0.8\columnwidth]{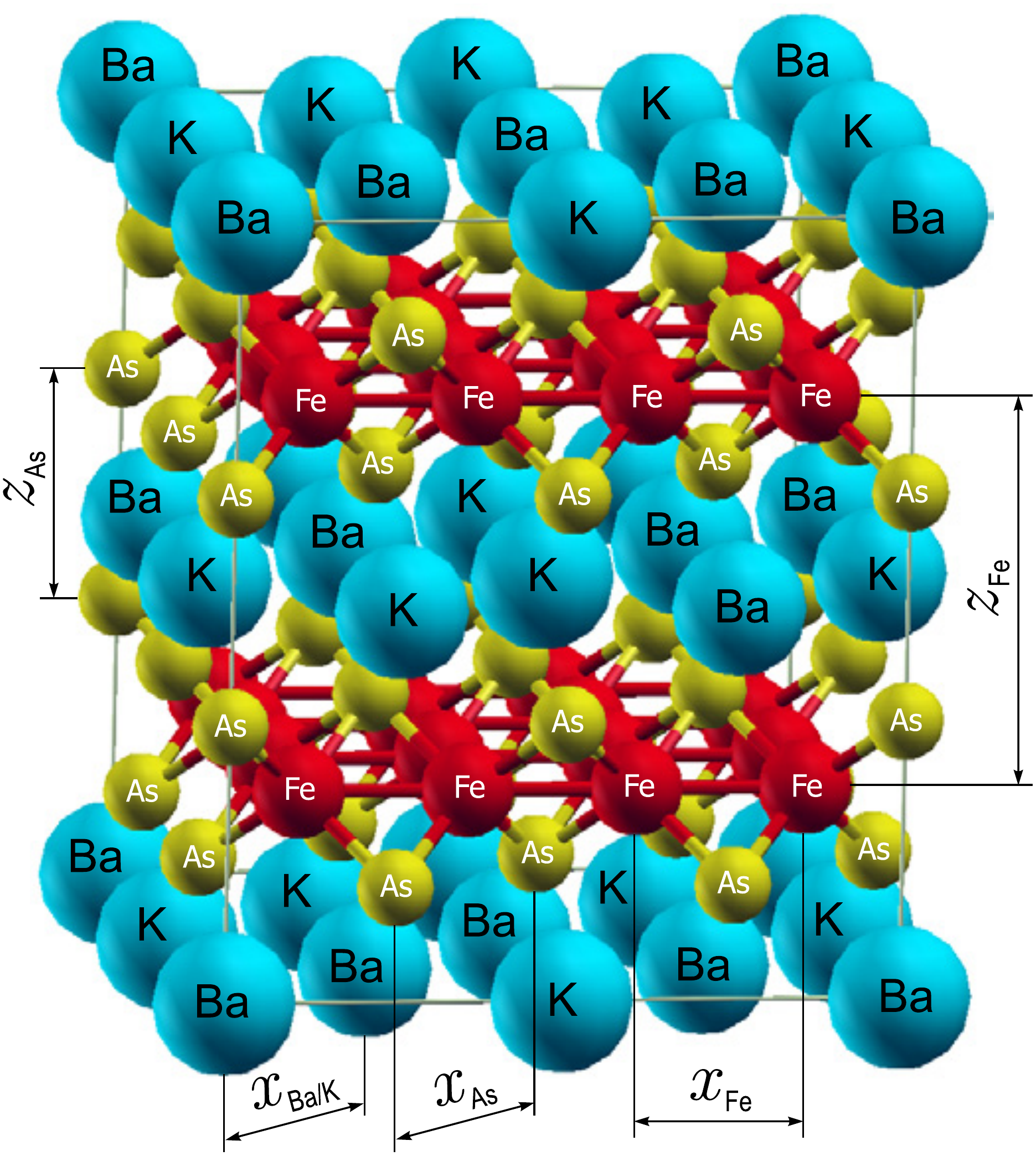}
\caption{The $2\sqrt{2}\,a\times2\sqrt{2}\,b\times c$ supercell with 50\% of the Ba atoms randomly substituted by the K dopants that we used in our density functional calculations. The calculated statistical distributions of the five interatomic distances, which are marked in the figure, are presented in Fig.\,\ref{fig:histograms}.\vspace{-3pt}}
\label{fig:unitcell}
\end{figure}

The microstrain distribution represents the statistics of the deviations ${\scriptstyle\Delta}d$ of the interplanar spacings from their average values, normalized by the average spacings $d$, i.e. of the strain $\varepsilon={\scriptstyle\Delta}d/d$, over the investigated specimen as a function of the crystallographic direction. Tensor surfaces representing the squared FWHM of the anisotropic microstrain distribution $B_\varepsilon^2$ along different crystallographic directions are shown as insets in Fig.\,\ref{fig:xray} (a) and (b), whereas panel (c) shows the $x$-$z$ [tetragonal (\textit{ac}) plane] and $x$-\kern-0.5pt$y$ [tetragonal (\textit{ab}) plane] cross-sections of both surfaces. The largest microstrains of the crystalline lattice both at 300\,K and at 15\,K are found in the $c$-direction ($|B_\varepsilon|_\perp=0.9$\% and 1.1\%, respectively$)$ as compared to the average in-plane values of $|B_\varepsilon|_\parallel=0.65$\% and 0.82\%. The flowerlike shape of the $x$-$z$ cross-section indicates a negative correlation between the in-plane ($hk0$) and the out-of-plane ($00l$) directions, which agrees with the opposite changes of the $a$ and $c$ lattice constants upon the variation of doping \cite{RotterPangerl08}.

\begin{figure}[t]
\includegraphics[width=\columnwidth]{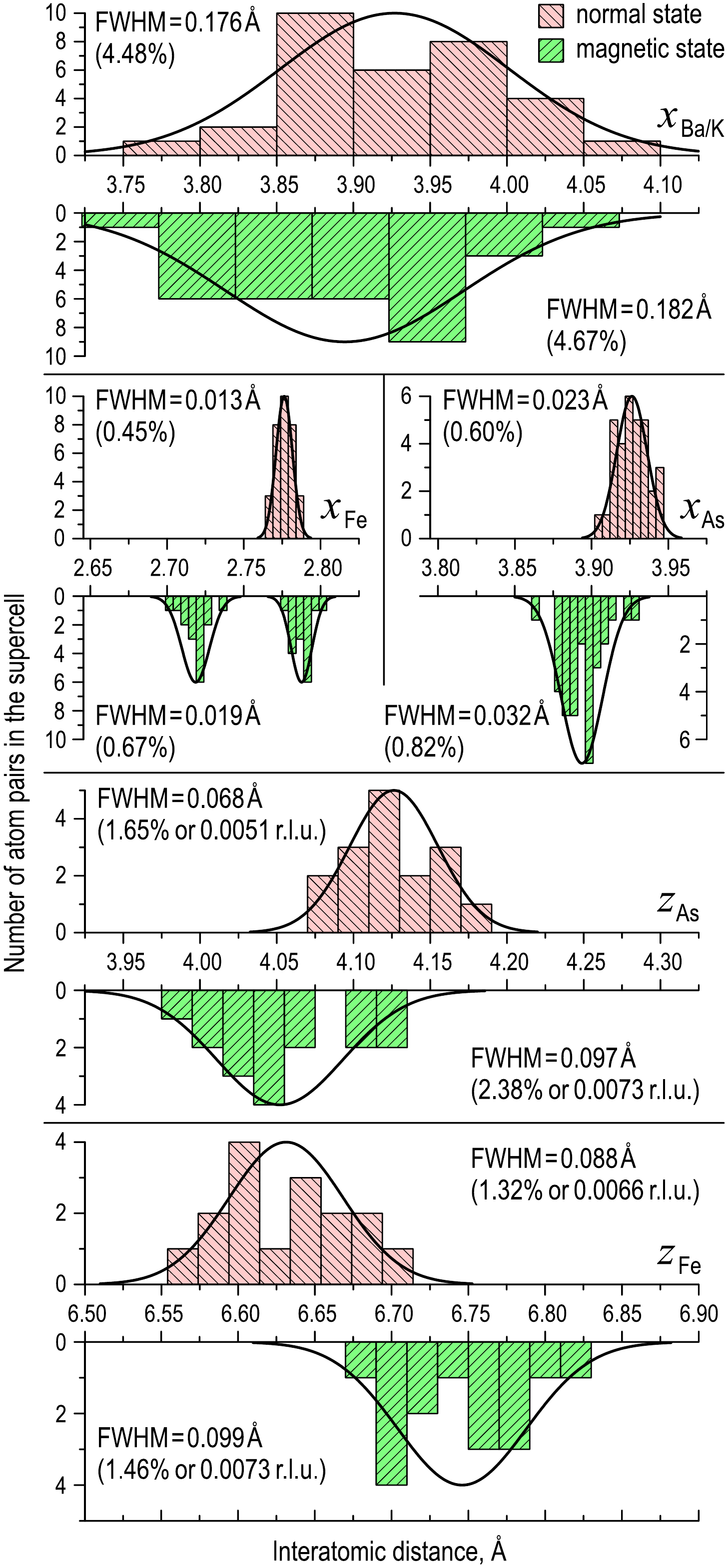}
\caption{Histograms of the calculated interatomic distances for the supercell as defined in Fig.~\ref{fig:unitcell}. The plots at the top of each panel are a result of the normal-state (non-magnetic) calculation, whereas the up-side-down plots below represent the magnetic ground state. The solid lines are fits to a normal distribution.\vspace{-12pt}}
\label{fig:histograms}
\end{figure}

The low-temperature increase of the microstrain amounts to $\sim$\,20\% relative to the corresponding values at room temperature both in the $c$-direction and in-plane. In other words, to a good approximation the two tensor surfaces are geometrically similar to each other, which would not be expected in the case of a weak orthorhombic distortion, as it should instead broaden only the in-plane peaks. Moreover, at both temperatures no considerable in-plane anisotropy is observed [i.e.\,anisotropy in the $x$-$y$ plane, see Fig.\,2(c)], which would be a sign for the onset of an orthorhombic phase transition, e.g. for an incomplete orthorhombic reflection splitting. Such an anisotropy (cf. also Ref.\,\onlinecite{Leineweber07}, Fig.\,1) characteristically precedes tetragonal-to-orthorhombic structural phase transitions in Pb$_3$O$_4$ \cite{DinnebierCarlson03} and La$_2$NiO$_4$ \cite{RodriguezCarvajal91}.

This lets us conclude that the origin of the microstrain at both temperatures is not related to a macroscopic structural transition to orthorhombic symmetry, but rather should be attributed to an increase of the microscopic distortions of the lattice. The microstrain distribution quantitatively represents the response of the lattice to structural defects, such as chemical inhomogeneities or dislocations, which are unavoidable in any real material. Therefore an increase of the microstrain below the magnetic transition can either indicate that the lattice becomes softer, i.e.~increases its response to the local stresses upon entering the AFM state, or that the local stresses themselves increase, causing a proportional increase of the microstrain.

In the studied compound, both mechanisms could be important. On the one hand, in the case of lattice softening, one would expect its direct influence on the phonon mode frequencies. Indeed, such an effect has been reported in the phonon spectra of two similar 122-compounds: polycrystalline Sr$_{0.6}$K$_{0.4}$Fe$_2$As$_2$ and Ca$_{0.6}$Na$_{0.4}$Fe$_2$As$_2$ \cite{MittalSu08}. There, softening of phonon modes below 10\,meV has been observed by inelastic neutron scattering upon cooling from 300\,K to 140\,K, despite the decrease of the unit cell volume at low temperature. More recently, softening and narrowing of several phonon modes below the spin density wave transition was also observed by Raman scattering in underdoped Sr$_{1-x}$K$_x$Fe$_2$As$_2$ and in the parent BaFe$_2$As$_2$ single crystals \cite{RahlenbeckSun09}. On the other hand, the phase-separated coexistence of AFM and paramagnetic phases in this material \cite{ParkInosov09} and the presence of twin AFM domain boundaries \cite{MazinJohannes09} should lead to an increase of local stresses below $T_{\rm m}$ due to the magnetic anisotropy of individual AFM domains.

To quantify the relative role of these two possible causes of the increased microstrain, we present here an estimation of the lattice softening across the magnetic transition based on our density functional calculations and show that it is comparable in magnitude and therefore could possibly provide a considerable contribution to the additional microstrain observed in the XRPD measurements.

\vspace{-5pt}\section{Density functional calculations}\vspace{-5pt}

Density functional calculations were performed using the projector augmented-wave \cite{Blochl94, KresseJoubert99} method in the framework of the generalized gradient approximation \cite{PerdewBurke97, PerdewBurke96}. We have chosen a large $2\sqrt{2}\,a\times2\sqrt{2}\,b\times c$ supercell, where 50\% of the Ba atoms were substituted by K to model a random distribution of the dopants, as shown in Fig.\,\ref{fig:unitcell}. Using the Vienna \textit{ab initio} simulation package (VASP) plane wave code \cite{KresseHafner93, KresseHafner94, KresseFurthmueller96prb, KresseFurthmueller96cms, KresseFurthmueller99}, we have carried out the crystal structure optimization of the cell parameters and all ionic positions within the supercell to determine their displacements from the high-symmetry positions due to the introduced chemical disorder. The unit cell volume was fixed during the course of structure optimization to the experimental room-temperature value for the parent compound, as derived from $a$\,=\,$b$\,=\,3.9625\,\AA, $c$\,=\,13.0168\,\AA\ \cite{RotterTegel08PRB}, which also agrees well with our room-temperature value measured for BKFA. The cutoff energy of the plane-wave expansion was 367 eV, and the Brillouin zone sampling mesh was 4\,$\times$\,4\,$\times$\,4 with its origin at the $\Gamma$ point. In the final optimized geometry, no forces on the atoms exceeded 0.01\,eV/\AA.

First, a non-magnetic calculation was performed, which represents the high-temperature (normal) state. As expected, it revealed no deviations from the tetragonal symmetry, and resulted in the lattice parameters of $a$\,=\,$b$\,=\,3.927\,\AA\ and $c$\,=\,13.258\,\AA, which are reasonably consistent with the results of the XRPD structure refinement discussed above.

The results of the crystal structure optimization are presented in Fig.\,\ref{fig:histograms} (top plot in each panel), which shows histograms of the five interatomic distances, as defined in Fig.\,\ref{fig:unitcell}, fitted to a Gaussian distribution (solid lines). Naturally, the largest atomic displacements due to potassium substitution are observed in the Ba/K plane itself, where the distances $x_{\rm Ba/K}$ between the neighboring Ba/K atoms vary by 0.176\,\AA\ or 4.48\% (0.0448~r.\,l.{\kern0.5pt}u., where r.\,l.{\kern0.5pt}u. stands for relative lattice units), as estimated by the FWHM of the distribution. Within the FeAs block of layers, the out-of-plane atomic displacements (the buckling of the As and Fe planes) are the largest, and amount to 0.068\,\AA\ (1.65\% or 0.0051~r.\,l.{\kern0.5pt}u.) for the As layer and 0.088\,\AA\ (1.32\% or 0.0066~r.\,l.{\kern0.5pt}u.) for the Fe layer. In-plane distortions are notably smaller: 0.023\,\AA\ (0.60\%) and 0.013\,\AA\ (0.45\%), respectively.

Finally, we performed a spin-polarized calculation for the low-temperature striped AFM state. Collinear magnetic moments were self-consistently determined within the calculation. The corresponding histograms are shown in the same figure at the bottom of each panel. The most noticeable effect is the splitting of the Fe-Fe nearest neighbor distance, $x_\text{Fe}$, which indicates the tendency of the system towards an orthorhombic distortion despite the presence of the dopants. As already mentioned above, such a transition is however suppressed macroscopically in the sample due to the presence of twin AFM domains and phase separation. In addition, one sees that the nearest-neighbor Fe-Fe inter-layer distance $z_\text{Fe}$ increases, while that of As, $z_\text{As}$, decreases, which corresponds to the stretching of the Fe-As tetrahedra.

Of more relevance for the present paper is the small but not negligible increase in the width of the distribution for every interatomic distance, as compared to the normal state, which we associate with the sought lattice softening effect. The variation of interatomic distances in the Ba/K plane increases by 3\% to 0.182\,\AA\ (4.67\% or 0.0467~r.\,l.{\kern0.5pt}u.). The buckling of the As and Fe planes increases to 0.097\,\AA\ (2.38\% or 0.0073~r.\,l.{\kern0.5pt}u.) and 0.099\,\AA\ (1.46\% or 0.0073~r.\,l.{\kern0.5pt}u.), respectively, which represents an increase by 43\% and 12\% relative to the corresponding normal-state values. In-plane distortions increase to 0.032\,\AA\ (0.82\%) for the As layer and 0.019\,\AA\ (0.67\%) for the Fe layer ($\sim$\,40\% of relative increase in the width).

The observed changes in FWHM of the distributions between the AFM and the normal states are statistically significant and are observed consistently for all five considered interatomic distances. For the FeAs block of layers, they amount to $\sim$0.002~r.\,l.{\kern0.5pt}u. on average both in- and out-of-plane. This is comparable with the increase of the microstrain $|B_\varepsilon|_\perp(15\,K)-|B_\varepsilon|_\perp(300\,K)\!=0.002$ and $|B_\varepsilon|_\parallel(15\,K)-|B_\varepsilon|_\parallel(300\,K)\!=0.0017$ observed in our XRPD experiment. We can therefore conclude that the softening of the lattice associated with the SDW transition provides a major contribution to the observed effect. An additional contribution from the increase of the local stresses at the AFM domain boundaries due to weak local distortions within each domain can not be excluded, however.

\vspace{-5pt}\section{Summary and discussion}\vspace{-5pt}

We have presented an example of an iron pnictide superconductor, which does not break its tetragonal crystal symmetry macroscopically upon entering the magnetically ordered state. This conclusion is based on x-ray powder diffraction measurements with subsequent analysis of the microstrain anisotropy. Instead, we have observed a low-temperature increase of the microstrain proportionally in all crystallographic directions, which has a magnetic origin and mostly originates from the softening of the crystal lattice below the SDW phase transition.
A detailed ana\-ly\-sis of the lattice structure in the presence of randomly distributed dopant atoms has been presented both in the normal and AFM states, confirming this conclusion. This does not exclude that a weak orthorhombic distortion possibly happens on a microscopic scale within each AFM domain, leading to an increase of the local stresses at the domain boundaries, whereas the mesoscopic electronic phase separation \cite{ParkInosov09} suppresses the breakdown of the tetragonal symmetry on lateral scales larger than the typical size of the AFM domains. It can be argued that the observed magnetic state of the lattice represents a crossover between the well-developed orthorhombic phase at low doping levels and the normal tetragonal phase typical for the overdoped region of the phase diagram, where no magnetic transition occurs even at lowest temperatures.

\vspace{-5pt}\section*{Acknowledgments}\vspace{-5pt}

The experimental work was performed at the \textit{Morpheus} diffractometer and RITA-II spectrometer, both at the Swiss spallation source SINQ, Paul Scherrer Institut (PSI), Villigen, Switzerland, and the X16C beamline at the National Synchrotron Light Source, Brookhaven National Laboratory, USA. We acknowledge financial support from DFG in the consortium FOR538, as well as from the Bundesministerium für Bildung und Forschung (BMBF), and the Fonds der Chemischen Industrie (FCI). Special thanks to A.~Yaresko and O.~Khvostikova for helpful discussions.

\bibliographystyle{phpf}\bibliography{Orthorhombicity}

\end{document}